\documentclass[lettersize,journal]{IEEEtran}
\IEEEoverridecommandlockouts

\usepackage{amsmath,amssymb,amsthm}
\usepackage{cite}
\usepackage{graphicx}
\usepackage[caption=false,font=normalsize,labelfont=sf,textfont=sf]{subfig}\usepackage{booktabs}
\usepackage[hidelinks]{hyperref}
\usepackage{microtype}
\usepackage[T1]{fontenc}
\usepackage{orcidlink}

\newtheorem{theorem}{Theorem}
\newtheorem{lemma}[theorem]{Lemma}

\title{Single-Connection Mixed-Criticality Transport with CATS: %
  Bounded Guarantees, Three Structural Limits, and a QUIC Escape}

\author{%
  \IEEEauthorblockN{Syed Muhammad Aqdas Rizvi~\orcidlink{0009-0004-1491-4839} \\}
  \IEEEauthorblockA{Independent Researcher\\
  Alumnus, Lahore University of Management Sciences (LUMS) \\
  Karachi, Pakistan\\
  \href{mailto:s.muhammadaqdasrizvi@gmail.com}{s.muhammadaqdasrizvi@gmail.com} | \href{mailto:25100166@lums.edu.pk}{25100166@lums.edu.pk}}
}

\begin{document}

\maketitle

\begin{abstract}
Satellite terminals, industrial telemetry-and-control, embedded systems,
tactical networks, and other constrained links often multiplex a small,
latency-critical message class with bulk traffic over a single commodity
transport connection.
A single FIFO connection can starve the critical class under load.
The obvious alternative, opening parallel connections, costs an additional
five-tuple (often blocked by carrier-grade NAT, port budgets, and operator
policy) and is not always available; when the critical class is light, two
connections can also be bandwidth-fair only in aggregate rather than
single-flow fair.
We present CATS (Conductor-driven Asymmetric Transport Scheme), a
sender-side, receiver-transparent transport-layer priority scheme over
commodity TCP: a \emph{Conductor} assigns each message a priority class and
\emph{just-in-time} sequence numbers, governed by a credit-based shaper.
CATS provides the one combination its alternatives cannot: deterministic
non-starvation together with single-flow fairness, plus a provable bounded
per-class delay.

We then show that CATS-over-TCP is not a tail-latency mechanism, and why.
Three structural barriers bound single-connection in-band priority at
three layers: the in-order sequence space (head-of-line blocking), the shared
congestion window (cross-class coupling), and the per-flow granularity of
network QoS (in-band priority is invisible to it).
The same barriers explain why fair-queuing and even the modern low-latency
standard L4S not only fail to help a single connection but actively make
it worse, by granting a parallel-connection competitor per-flow isolation
while leaving in-band priority unserved, and why two parallel
connections reduce the latency tail by 4--8$\times$ at the cost of an
additional flow.
We give CATS-over-QUIC as the principled escape: independent sequence spaces
with per-stream isolation under aggregate-coupled congestion control
self-isolate at the endpoint, attaining the guarantees on one fair flow.
A proof-of-concept shows unreliable datagrams reach a flat critical-class
tail ($\approx$27\,ms p99.9 vs.\ $\approx$307\,ms over TCP) by never
retransmitting a lost message, so it never waits on the bulk stream's
recovery, while reliable streams still need an
aggregate-coupled congestion controller to bound what remains; we frame
that controller, a gap the QUIC standards leave open, as a research
program.
An ns-3 evaluation across loss, contention, and handover regimes supports
every claim, including the negative ones.
\end{abstract}


\section{Introduction}
\label{sec:intro}

Two very different kinds of traffic routinely share one transport
connection: a small, latency-critical class (control commands, coordination messages, model-routing decisions) interleaved
with bulk transfers such as logs, model weights, sensor dumps, or media.
This pattern is the norm in settings where a second connection is costly
or unavailable: satellite and non-terrestrial terminals behind carrier-grade
NAT, embedded and industrial endpoints with tight socket and port budgets,
and existing single-socket applications that cannot be re-architected.
In these settings the critical class and the bulk class are not separable
into independent network flows; they share one sequence space, one congestion
controller, and one path.

A single first-in-first-out connection serves this mix badly.
Under load the critical class queues behind the bulk backlog and starves:
in our experiments a FIFO connection delivers 0 of 300 critical messages
within the measurement window under sustained contention, because the head of
the send buffer is perpetually bulk data.
The usual remedy is a parallel connection for the critical class, and it
works, but at a cost that is easy to overlook: \emph{N} parallel
connections require \emph{N} separate five-tuples, which are blocked by
carrier-grade NAT, tight socket and port budgets, single-socket application
constraints, and operator policy, and may be unavailable in practice.
When the critical class is light, the parallel-connection setup can still look
bandwidth-fair in aggregate; in our measurements, the Jain fairness
index~\cite{jain_fairness} is
$\approx 0.67$ for two connections versus $\approx$1.0 for one, reflecting a thin-flow
accounting effect rather than a fundamental improvement in end-to-end fairness.
Two connections are bandwidth-fair when the critical class is light; CATS is
single-flow-fair unconditionally, regardless of class bandwidths.

This paper asks what a single connection can guarantee for
mixed-criticality traffic, and what it fundamentally cannot.

\textbf{CATS and its guarantees.}
We present CATS (Conductor-driven Asymmetric Transport Scheme), a sender-side, receiver-transparent transport-layer priority
scheme over commodity TCP.
A Conductor tags each application message with a priority class and
assigns sequence numbers just in time, so a newly-arriving
high-priority message takes the next transmission slot rather than queuing
behind buffered bulk bytes.
A two-threshold credit-based shaper governs how classes share the connection.
CATS needs no receiver changes, no new congestion-control algorithm, and no
kernel modifications.
On one connection it provides the combination no alternative offers at once:
deterministic non-starvation (the critical class always makes
progress, 300/300 delivered where FIFO delivers 0/300),
single-flow fairness (exactly one TCP-fair flow), and a
provable bounded per-class delay (Theorem~\ref{thm:bound}).
Among single-flow-fair options, CATS is the only non-starving one.
The core concept of CATS originally appeared in~\cite{cats2025icic}, with 
the further theory detailed in the extended text~\cite{cats2026arxiv}; the just-in-time architecture, the three-barrier structural result, and the QUIC escape developed here are completely new.

\textbf{CATS's limits.}
CATS-over-TCP is not a tail-latency mechanism, and the
reason is structural, not an artifact of our design.
Under cross-traffic, two parallel connections win the critical-class tail by
4--8$\times$, and they win it under handover as well.
We trace this to three structural barriers that bound any
single-connection in-band priority scheme, each at a different layer:

\begin{itemize}
  \item \textbf{(B1) Sequence space.}
    One in-order sequence space (in TCP, one byte stream): a low-priority loss head-of-line-blocks the
    critical class for at least a round trip, no matter how the sender
    schedules, because in-order delivery is enforced at the receiver.
  \item \textbf{(B2) Congestion control.}
    One shared congestion window: a bulk-triggered loss contracts it and so
    delays the critical class too, by a quantifiable coupling term
    $\Delta_{cc} = 2\,\mathrm{RTT}(1-\beta)$.
  \item \textbf{(B3) Network-queue granularity.}
    Network QoS (fair-queuing, DSCP AQM, L4S): this schedules at
    flow granularity; a single connection is one five-tuple, so its
    in-band priority is invisible to the network, which therefore cannot
    isolate the critical class even when asked.
\end{itemize}

These bite at three layers and explain the three distinct findings of the loss
tail (B1), the contention coupling (B2), and the counter-intuitive one (B3):
adding fair-queuing at the bottleneck makes a single connection's position
worse, because it grants a parallel-connection competitor per-flow
isolation while leaving in-band priority unserved (and L4S, classifying by
per-flow marking, is no different).
The barriers are properties of a single TCP connection.
Two connections clear B1, B2, and B3 but at the cost of N five-tuples;
\textbf{CATS-over-QUIC} clears B1 with independent sequence spaces and bounds B2 with
per-stream isolation under aggregate-coupled congestion control: a priority
reservation preserves the critical class's share of the window once a bulk
loss cuts it, though it cannot prevent the cut itself, since that cut is what
keeps the connection one fair flow. This self-isolation at the endpoint means
it no longer needs the network, thus rendering B3 moot from above, all on
one fair flow.
A proof-of-concept confirms the structural bypass: unreliable datagrams reach a flat
critical-class tail under loss ($\approx$27\,ms p99.9 vs.\ $\approx$307\,ms over TCP
(per-seed-median p99.9), CCA-independent) because a lost datagram is never
retransmitted and so never waits on the bulk stream's recovery,
while reliable streams still need the aggregate-coupled controller to bound what remains,
pinning the precise requirement: per-stream isolation under aggregate-coupled congestion control.

\textbf{On scope.}
We report the regimes where two connections beat CATS-over-TCP rather than around them:
CATS-over-TCP's value is non-starvation, fairness, and a bounded floor on one
connection, competitive on latency only as the dominant flow, and a
guarantor of completeness, not of the tail.

\textbf{Contributions:}
\begin{itemize}
  \item \textbf{CATS-over-TCP}, a sender-side, receiver-transparent priority scheme
    over commodity TCP, providing deterministic non-starvation with single-flow fairness, together with a provable bound on per-class delay, via a credit-based-shaper model and a just-in-time sequencing realization
    (Sections~\ref{sec:design},~\ref{sec:theory}).
  \item \textbf{A three-barrier structural result} across the sequence space
    (B1), congestion window (B2), and network-queue granularity (B3), bounding
    single-connection in-band priority, with B2 quantifying the contention
    coupling and B3 explaining why fair-queuing and L4S cannot help a
    single connection (Section~\ref{sec:theory}).
  \item \textbf{CATS-over-QUIC}, the principled escape via endpoint
    self-isolation (independent sequence spaces + per-stream-isolated aggregate CC), with a
    proof-of-concept validating the structural bypass and identifying
    per-stream isolation under aggregate-coupled CC as the precise requirement
    (Section~\ref{sec:quic}).
  \item \textbf{A multi-regime evaluation} (ns-3 across loss,
    contention, and handover; plus the QUIC PoC) delineating exactly where
    single-connection priority helps and where it does not
    (Sections~\ref{sec:eval},~\ref{sec:quic}).
\end{itemize}

Section~\ref{sec:motivation} scopes the use-case envelope;
Section~\ref{sec:design} presents the CATS design;
Section~\ref{sec:theory} develops the per-class delay bound and three
structural barriers;
Section~\ref{sec:eval} evaluates CATS-over-TCP across five regimes;
Section~\ref{sec:quic} presents CATS-over-QUIC and the proof-of-concept;
Sections~\ref{sec:related}--\ref{sec:conclusion} discuss related work,
limitations, and conclusions.

\section{Motivation and Use-Case Envelope}
\label{sec:motivation}

Single-connection in-band priority is the relevant tool only in a narrow
envelope, and that envelope shapes every design choice and every
experiment that follows.

\textbf{The traffic:}
we consider two classes sharing one connection: a critical class P0
of small, sporadic, latency-sensitive messages (control, coordination,
acknowledgement, routing decisions on the order of a few hundred bytes
each), and a bulk class P4 of throughput-oriented transfers that is
essentially always backlogged (logs, weights, media, sensor data).
In our experiments P0 is 200-byte messages generated at 10\,msg/s;
P4 is a 50\,MB perpetual backlog.
The classes are application-meaningful but not network-separable: the
application has one socket, and the two classes are interleaved bytes on
one sequence space.

\textbf{Why one connection is the binding constraint:}
opening a second connection for the critical class is the textbook answer,
and where it is feasible and its extra share acceptable it is hard to beat.
But it is frequently not feasible: carrier-grade NAT that limits or
rewrites flows, tight socket and port budgets on embedded endpoints,
single-socket applications that cannot be re-architected, or operator policy
that permits one flow. Even where feasible, it costs N five-tuples (blocked by carrier-grade NAT,
port budgets, and operator policy) and is bandwidth-fair only when the
critical class is light (a heavier critical class would consume a second share).
Single-connection priority is the right tool when one must, or
should, present a single flow to the network.

\textbf{Why constrained links sharpen the problem:}
low-Earth-orbit (LEO) and non-terrestrial networks, cellular backhaul, and
constrained IoT links are where this matters most: they combine meaningful
loss, elevated round-trip times, and periodic disruption,
which make the FIFO baseline fail and the extra-connection alternative most
costly.
For LEO, disruption is periodic: Starlink hands over every 15\,s, producing
latency spikes and loss bursts that confuse congestion control~\cite{leo_mobility};
our handover experiments use an explicit outage model aligned to this cadence.
CATS is orthogonal to the handover-aware congestion control this
community has converged on: those repair the controller's reaction to a
handover, while CATS arbitrates priority within a connection; the
two compose, and we treat them as complementary.

\textbf{The envelope, stated:}
CATS-over-TCP is worth deploying when (i)~the application multiplexes a
small critical class with bulk over (ii)~a single connection that cannot be
split into independent flows, on (iii)~a constrained link where loss and
disruption make FIFO starvation real, and where (iv)~presenting more than
one flow to the network is impossible or undesirable.
Outside this envelope (when a second connection is free and its extra
share is acceptable, or when the link is uncongested and lossless), the
problem is either better solved by parallel connections or not a problem
at all.
Everything below is evaluated against this envelope, and Section~\ref{sec:eval}
reports where even inside it the single connection is beaten on
latency.

\section{CATS Design}
\label{sec:design}

CATS-over-TCP adds priority to a single commodity-TCP connection entirely at the
sender, with an unmodified receiver and an unmodified congestion controller.
It has three elements: the Conductor, which assigns priority and
arbitrates the connection; a credit-based shaper, which governs how
classes share transmission; and just-in-time sequencing, which
integrates the priority decision into the TCP send path.
Throughout this paper, CATS-over-TCP refers to this just-in-time (JIT)
integrated-sender design.
It supersedes the original interceptor/feeder architecture
of~\cite{cats2025icic}, which arbitrated priority in a shim above the
socket and therefore could not prevent the send-buffer queuing that JIT
eliminates, and we treat that earlier design as legacy throughout.

\textbf{The Conductor:}
the application submits messages tagged with a priority class.
The Conductor maintains a per-class queue and decides, at each transmission
opportunity, which class's data is sent next.
Because all classes ride one TCP connection, the receiver reassembles a
single ordered byte stream with no awareness of priority. CATS-over-TCP is invisible
on the wire beyond ordinary TCP and requires no receiver cooperation.

\textbf{Credit-based shaper and non-starvation:}
sending strictly in priority order would starve the bulk class and,
under adversarial arrival patterns, destabilize the connection.
CATS-over-TCP instead governs class arbitration with a credit-based shaper using two
thresholds (a hysteresis band): a class accrues credit over time and spends
it to transmit, and the two-threshold band bounds how long any class waits
before it is served.
This yields deterministic non-starvation: every class is served
within a bounded interval, so the critical class makes progress even against
a perpetually-backlogged bulk class (where FIFO delivers none of it),
and a provable bounded per-class delay developed in
Section~\ref{sec:theory}.

\textbf{Just-in-time sequencing:}
a subtle failure mode defeats naive sender-side priority: if the Conductor
selects a high-priority message but the TCP send buffer already holds a
backlog of bulk bytes with assigned sequence numbers, the high-priority
message still waits behind that backlog, because TCP transmits in sequence
number order.
CATS-over-TCP closes this gap by assigning TCP sequence numbers just in time:
at the moment a message is handed to the transmission path, not when it
is enqueued, so a newly-arriving critical message takes the next on-wire
slot rather than queuing behind the buffered bulk backlog.
This integrated-sender design removes the send-buffer latency floor
measured in Fig.~\ref{fig:e16}; the remaining floor is the fundamental
per-class delay bound of Section~\ref{sec:theory}.

\textbf{Optional mitigations:}
beyond the core JIT design, CATS-over-TCP admits optional add-ons, namely DSCP marking for
managed networks, message duplication/FEC, and packet-level redundancy.
These are separable from the core architecture and required by none of
its guarantees.
We evaluate them in Section~\ref{sec:eval} and find that only duplication/FEC
earns its complexity, and only at loss rates $\geq$2\%.
DSCP is inert without AQM-aware hardware and, by the network-queue barrier
(Section~\ref{sec:theory}), cannot help a single connection in any case.
We keep all three out of the core and treat them as deployment options.

\textbf{What CATS-over-TCP does not require, and does not change:}
CATS-over-TCP uses commodity TCP and is agnostic to the congestion-control algorithm:
it runs unchanged over BBR and CUBIC, and composes with specialized
controllers (e.g., LEO handover-aware controllers).
These properties are what make CATS-over-TCP deployable today; as the next
section shows, they also bound what it can achieve.

\section{Theory}
\label{sec:theory}

This section develops what CATS guarantees on a single connection, and
what single-connection in-band priority can achieve at all.
The guarantees and the barriers are two sides of one fact: a single commodity
connection is one in-order sequence space, one congestion window, and one
network-visible flow.

We name the three barriers \textbf{B1} (sequence space), \textbf{B2}
(congestion control), and \textbf{B3} (network queue), and we are deliberate
about their epistemic status: B1 is a deterministic lower bound (Theorem~\ref{thm:b1}),
B2 is a quantified coupling bound (Lemma~\ref{lem:b2}), and B3 is an architectural
barrier witnessed empirically, not a deductive theorem.

\subsection{Model and Guarantees}
\label{sec:theory:model}

\textbf{Model:}
one TCP connection carries $K$ priority classes.
The Conductor assigns each class a credit rate $r_i$, an upper threshold
$H_i$, and a lower threshold $L_i$; class $i$ transmits when its credit
balance is positive.
Let $R$ be the bottleneck link rate, $d_{\mathrm{prop}}$ the one-way
propagation delay, $B_c$ the maximum per-class buffer occupancy,
MSS the maximum segment size, and $b_0$ the per-class lower credit threshold.
Under a Gilbert-Elliott loss model with burst parameter $p$,
let $L_{\mathrm{loss}}$ denote the expected extra delay per loss event.

\textbf{Bounded per-class delay:}
\vspace{-5.5pt}
\begin{theorem}[Per-class delay bound]
\label{thm:bound}
Under the CATS credit-based shaper, the worst-case delivery time of
a critical-class (P0) message is bounded by
\begin{equation}
  D_0 \;\leq\;
    \frac{B_c + \mathrm{MSS} + b_0}{R}
    + d_{\mathrm{prop}}
    + L_{\mathrm{loss}}
    + \Delta_{cc},
\end{equation}
where
$\Delta_{cc} = 2\,\mathrm{RTT}(1-\beta)$
is the congestion-coupling term
($\beta$ is the CC's multiplicative decrease factor;
BBR~\cite{bbr}: $\beta \approx 0.75$, CUBIC~\cite{rfc9438}: $\beta = 0.7$, NewReno~\cite{rfc6582}: $\beta = 0.5$).
\end{theorem}
At the ns-3 topology (5\,Mbps, OWD = 25\,ms, BBR), the loss-only floor
$D_{0,\mathrm{loss}}$ (setting $\Delta_{cc} = 0$) $\approx$27\,ms;
the full bound with BBR coupling
$D_{0,cc} \approx 53$\,ms ($\Delta_{cc} = 25$\,ms for RTT = 50\,ms,
$\beta = 0.75$).
Section~\ref{sec:eval:single} shows the measured JIT floor agrees with
$D_{0,\mathrm{loss}}$ (Fig.~\ref{fig:e16}).

\textbf{Non-starvation, conditioned on connection liveness:}
the two-threshold band guarantees that every class is served within a
bounded interval whenever the connection is making progress.
This is the formal counterpart of the empirical 300/300-vs-0/300 result.
\textbf{Scope:} non-starvation is a scheduling-layer guarantee conditional
on the connection having transmission opportunities.
It does not override a connection-level congestion collapse (B2's
unbounded regime), in which the window is driven to zero and no class can
be served.
Section~\ref{sec:eval:handover} exhibits this case at high round-trip times.
On one connection that is making progress, CATS-over-TCP is the only single-sequence-space
scheme that is both single-flow-fair and non-starving.

\subsection{Three Structural Barriers on a Single Connection}
\label{sec:theory:barriers}

\textbf{B1---Sequence-space barrier (Theorem~2).}
\vspace{-5.5pt}
\begin{theorem}[Sequence-space barrier]
\label{thm:b1}
On a single TCP connection (or any single in-order stream),
critical-class message $m$ is delivered no earlier than the last ACK
of every byte $b$ with $\mathit{seq}(b) < \mathit{seq}(m)$.
If any such byte $b$ is lost, P0 delivery is delayed by at least
one retransmission round:
\[
  \mathit{delivery\_time}(m)
    \;\geq\;
  \mathit{recv\_time}(\text{last byte before }m) + 1\,\mathrm{RTT}.
\]
\end{theorem}
\noindent
The consequence is that JIT, or any other sender-side scheduler,
cannot avoid B1: the sender controls transmission order, but not
the receiver's in-order delivery.
\textbf{Escape:} independent sequence spaces (separate streams
or unreliable datagrams), which only a multi-stream transport (QUIC)
or multiple connections provide.

\textbf{B2---Congestion-control barrier (Lemma~3).}
\vspace{-5.5pt}
\begin{lemma}[CC-independence barrier]
\label{lem:b2}
On a single connection with shared congestion window, a congestion event
that reduces the window by factor $(1-\beta)$ imposes an additional delay
\[
  \Delta_{cc} = 2\,\mathrm{RTT}(1-\beta)
\]
on all priority classes simultaneously, including P0, regardless
of sequence-space isolation.
\end{lemma}
\noindent
Derivation: at steady state, $\mathit{cwnd}_{ss} \approx 2 \cdot R \cdot \mathrm{RTT}$;
a multiplicative decrease by $(1-\beta)$ creates a surplus of
$\mathit{cwnd}_{ss}(1-\beta)$ unacknowledged bytes; the time to drain
the surplus at rate $R$ is $2\,\mathrm{RTT}(1-\beta)$.

Two points matter.
First, B2 bites even when B1 is solved: independent sequence spaces remove
head-of-line blocking but still share one congestion controller, so a bulk
loss still throttles the critical class, which is why independent sequence
spaces are necessary but not sufficient.
Second, the nominal coupling $\Delta_{cc}$ does not bound the rare
congestion-collapse regime (window driven to near-zero), which is
effectively unbounded until recovery; Section~\ref{sec:eval:handover}
exhibits this at high RTT under handover, and the QUIC PoC exhibits it in
one of twenty seeds (Section~\ref{sec:quic:poc}).
\textbf{Escape:} per-stream isolation under a shared aggregate controller.

\textbf{B3---Network-queue barrier (architectural, witnessed empirically).}
\paragraph{Barrier B3 (Network-Queue Granularity)}
Network QoS schedulers (FqCoDel~\cite{rfc8290}, DSCP-aware AQM, class-based weighted
fair queuing, and the modern low-latency standard L4S
\cite{rfc9330,rfc9331,rfc9332}) classify and schedule at
flow (five-tuple) granularity.
A single connection, TCP or QUIC, presents exactly one five-tuple;
its in-band priority signal (a CATS-over-TCP tag, a QUIC stream identifier)
lives inside the payload (after encryption, in QUIC's case) and is
invisible to the network.
Therefore a single connection's critical class receives no
per-priority treatment from network-layer QoS, even when the operator
configures prioritization.
We state this as a structural barrier and support it two ways.
Empirically (Section~\ref{sec:eval:fqcodel}): introducing FqCoDel at the
bottleneck widens a two-connection competitor's advantage over
a single CATS-over-TCP connection.
Architecturally, L4S classifies by per-flow ECN marking, so the same
conclusion applies: B3 is a statement about all flow-granular network QoS.
FQ slot counts confirm the mechanism: CATS-over-TCP occupies $1{+}K$ slots
under $K$ background flows; two-conn occupies $2{+}K$
(Section~\ref{sec:eval:fqcodel}).
\textbf{Escape:} none at the network for a single flow; per-priority network
isolation requires distinct flow identifiers (multiple connections,
each an additional five-tuple).
The way out is not to make the network see priority, but to no longer need
it to (Section~\ref{sec:quic}).

\subsection{The Three Barriers Together, and the Escape}
\label{sec:theory:table}

The three barriers bound single-connection in-band priority at three
independent layers and cleanly separate the design space (Table~\ref{tab:design}).
FIFO on one connection fails B1 and B2 and starves the critical class.
B3 does not apply, since FIFO carries no priority signal to hide.
CATS-over-TCP provides non-starvation, single-flow fairness, and the bounded
floor, but remains limited by all three on latency, which is why, as
Section~\ref{sec:eval} reports, two connections beat it on the tail.
Two parallel connections clear B1, B2, and B3 but require N five-tuples instead of one.
CATS-over-QUIC clears B1 with independent sequence spaces and bounds B2
with per-stream isolation under aggregate-coupled congestion control: the
priority-weighted allocation of the aggregate window preserves the critical
class's share of the window once a bulk loss cuts it, but it does not prevent
the cut, because the cut is what keeps the connection one congestion-fair
flow. This achieves endpoint self-isolation on both counts;
once the endpoint self-isolates, it no longer needs the network to isolate
it, rendering B3 moot from above.
(QUIC streams are not network-visible flow identifiers,
one encrypted five-tuple; this is not B3 defeated by visibility,
but B3 made irrelevant.)
This is achieved on one connection (one flow's fair share).

\begin{table}[t]
\centering
\caption{Design space: single-connection mechanisms vs.\ structural barriers.
  \protect\\
  B1=seq-space; B2=CC-coupling; B3=network-queue;
  ``moot''=endpoint self-isolation renders network treatment irrelevant.}
\label{tab:design}
\begin{tabular}{@{}lcccc@{}}
\toprule
Scheme & B1 & B2 & B3 & Fair / Non-starving \\
\midrule
FIFO-TCP         & \textbf{No} & \textbf{No} & N/A & Yes / \textbf{No}  \\
CATS-over-TCP    & mitigated         & bounded        & \textbf{No}   & Yes / Yes          \\
Two-conn TCP     & Yes            & Yes            & Yes  & \textbf{No} / Yes  \\
CATS-over-QUIC   & Yes            & bounded$^*$    & moot & Yes / Yes          \\
\bottomrule
\vspace{2pt}
\end{tabular}
{\footnotesize For B1--B3: Yes clears the barrier; bounded, a
proven delay ceiling; mitigated, reduced with no ceiling; No fails;
N/A, no signal to isolate; moot, a signal exists but the network need
not see it. $^*$PoC preserves the critical
class's share of the post-cut window.}
\end{table}

CATS-over-QUIC is the only point in the design space that is
simultaneously single-connection, network-fair, non-starving, clearing B1
and bounding B2, and the most tail-robust design available
on one connection.
Section~\ref{sec:quic} instantiates it and validates the structural bypass.

\section{Evaluation: CATS over TCP}
\label{sec:eval}

We evaluate CATS-over-TCP across the regimes of its envelope (single-connection
latency, contention, managed queuing, handover, and loss) and report, in each,
both what CATS-over-TCP guarantees and where two connections beat it.
The throughline is the one the theory predicts: CATS-over-TCP is competitive on latency
only when it is the dominant flow, its value is non-starvation and fairness,
and the three barriers are visible in the data.

\subsection{Setup}
\label{sec:eval:setup}

All experiments use ns-3 (release 3.38) with a dumbbell topology:
5\,Mbps bottleneck, 25\,ms one-way propagation delay (50\,ms base RTT),
Gilbert-Elliott 0.5\% bursty loss (mean burst length 10 packets),
and shallow-buffer BBR or deep-buffer CUBIC.
P0 is 200\,B messages at 10\,msg/s (300 messages total per run);
P4 is a 50\,MB perpetual backlog.
Schemes compared: CATS-over-TCP (the JIT-only design), two independent connections
(``two-conn''), single FIFO (``FIFO''), and CATS-legacy (interceptor/feeder,
for the JIT-floor comparison only).
Seed counts: 30 for the single-connection latency experiment, 20 for the
fairness and FqCoDel experiments, 40 for handover; 20 for the QUIC PoC.

\textbf{Reporting convention (survivorship bias):}
a seed with incomplete delivery contains only its early, fast messages;
the slow tail is absent, so including such seeds biases tail percentiles
low.
We therefore report all tail statistics (p99, p99.9) over
complete-delivery seeds only ($n_{\mathrm{recv}} = n_{\mathrm{sent}}$),
and report the incomplete-seed count separately as a first-class reliability
metric.
This correction matters materially only in the high-RTT/high-loss handover
regime; across the single-connection latency, fairness, FqCoDel, and QUIC
PoC experiments all seeds deliver completely.
Where corrections apply (Section~\ref{sec:eval:handover}), they make the
single connection's tail worse, reinforcing, not weakening, the
reported findings.

\subsection{Single-Connection Latency and the JIT Floor}
\label{sec:eval:single}

\begin{figure}[t]
  \centering
  \includegraphics[width=\columnwidth]{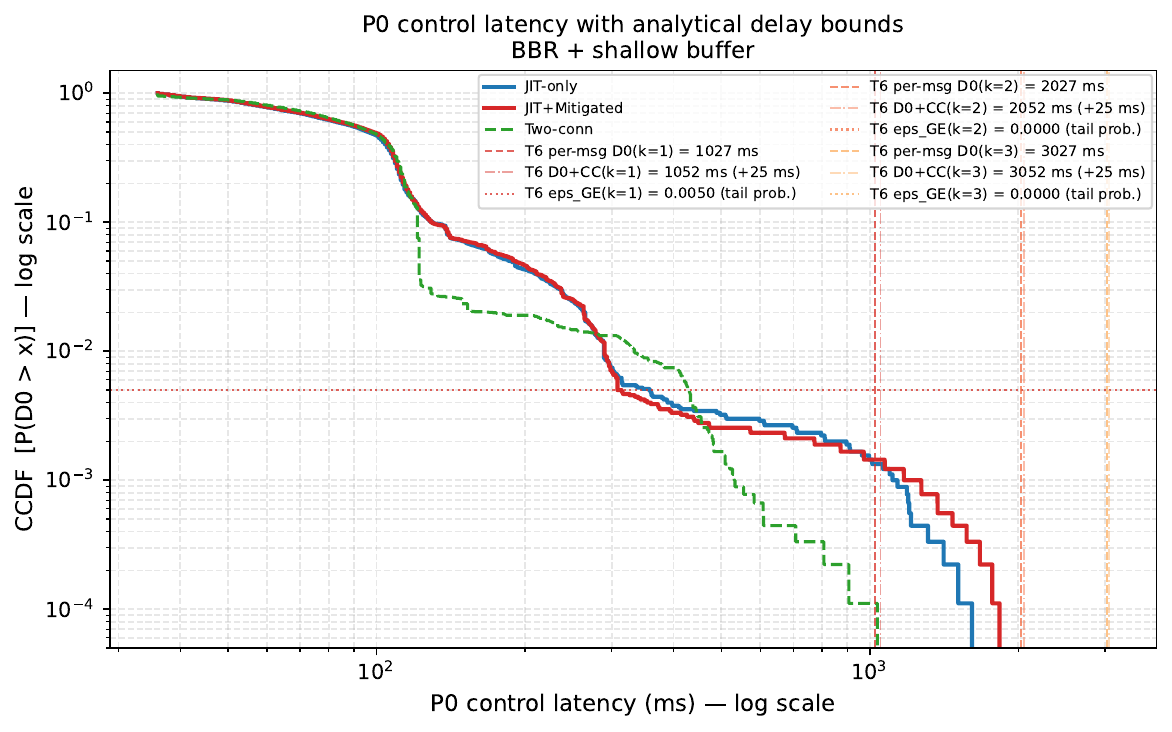}
  \caption{P0 critical-class latency CCDF (BBR/shallow buffer, 30 seeds).
    CATS-over-TCP collapses the legacy interceptor/feeder floor from
    $\approx$187\,ms to $\approx$94\,ms, matching the analytical bound
    $D_{0,\mathrm{loss}} \approx 27$\,ms (dashed).
    FIFO: 0/300 critical messages delivered (not shown).
    Shaded bands: cross-seed variance.}
  \label{fig:e16}
\end{figure}

On a single connection with no competing cross-traffic (CATS-over-TCP's home regime),
CATS-over-TCP is latency-competitive, and JIT sequencing is what makes it so.
The legacy interceptor/feeder design carries a send-buffer floor of
$\approx$187\,ms (p50, BBR/shallow) that JIT removes, collapsing the
critical-class median to $\approx$94\,ms (BBR; $\approx$78\,ms CUBIC);
the floor is now the analytical per-class bound $D_0$, not the buffer.

Against two connections in this regime, CATS-over-TCP is competitive at the
median and p99: per-seed-median p99 = 272\,ms vs.\ two-conn 314\,ms
(BBR/shallow; CATS-over-TCP wins both).
Two connections win only the deep tail:
pooled p99.9 = 531 vs.\ 1{,}112\,ms; pooled CVaR95 = 187 vs.\ 295\,ms
(complete seeds, $30{\times}300$ messages, BBR/shallow);
see Fig.~\ref{fig:e16} for the full CCDF.
The baseline starves: a single FIFO connection delivers 0 of 300 critical
messages, where CATS-over-TCP delivers 300/300.
The takeaway: uncontended and single-connection, CATS-over-TCP
is competitive on latency through p99, and it is the only single-flow-fair
scheme that is also non-starving;
two connections win only the deep tail (p99.9 and CVaR95).
The rest of this section shows that the moment we leave this regime, the
comparison changes sharply.

\subsection{Fairness and the Cost of Parallel Connections}
\label{sec:eval:fairness}

Under cross-traffic, two connections win the critical-class tail.
With $K{=}1$ background flow, two-conn p99 (240\,ms) is
$4.7{\times}$ better than CATS-over-TCP (1{,}116\,ms), measured against a Jain
fairness index of 0.671 vs.\ CATS-over-TCP's 0.996.
This low Jain index is a thin-flow accounting artifact: the thin P0
connection ($\approx$16\,kbps) underuses its nominal share, and the
background flow absorbs the slack; the two-connection side's aggregate takes
approximately one fair share.
Two connections are bandwidth-fair for this thin critical class; CATS-over-TCP is
unconditionally single-flow-fair regardless of class bandwidths.
The real cost of two connections is flow count: two five-tuples, blocked by
carrier-grade NAT, port budgets, and operator policy.
CATS-over-TCP, by construction one TCP-fair flow, stays fair while its
critical-class latency degrades under the same contention.
FIFO again delivers 0/300.
The tradeoff is that for a light critical class, both approaches are
bandwidth-fair, but two connections are faster on the tail and cost an
additional five-tuple; CATS-over-TCP is single-flow-fair unconditionally and can be
opened where only one connection is permitted.
The natural question is whether a fair-queued network closes the
gap further.
It does not.

\subsection{Managed Queuing Does Not Rescue a Single Connection (B3)}
\label{sec:eval:fqcodel}

\begin{figure}[t]
  \centering
  \includegraphics[width=\columnwidth]{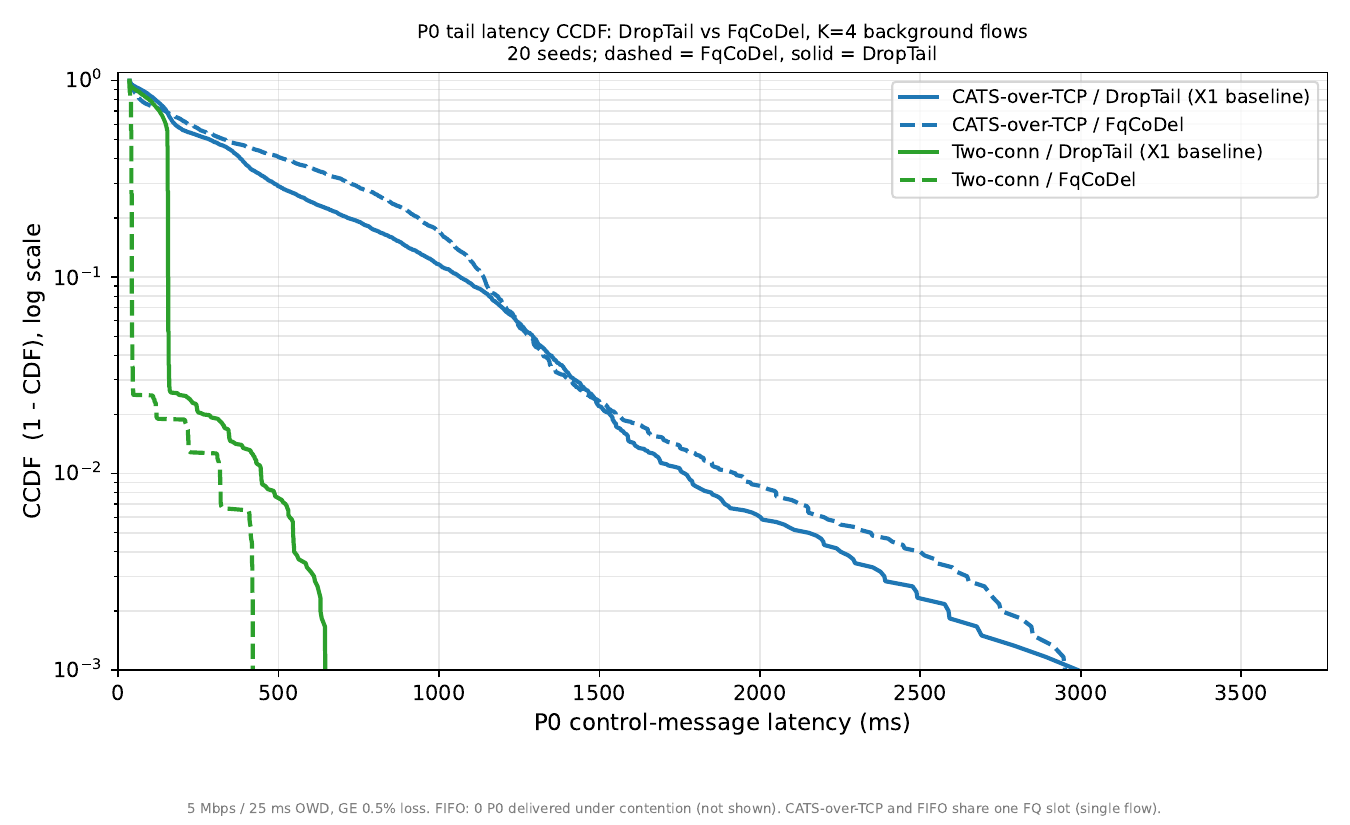}
  \caption{P0 latency CCDF with DropTail vs.\ FqCoDel ($K{=}4$
    background flows, 20 seeds).
    FqCoDel widens two-conn's lead (p50: 154\,ms~$\to$~42\,ms, $-73$\%)
    while CATS-over-TCP's p50 is unchanged or slightly worse
    (280\,ms~$\to$~333\,ms, $+19$\%).
    Two-conn Jain = 0.838 under both disciplines (unchanged Jain; thin-flow artifact, not a bandwidth grab).
    FIFO: 0/300 delivered.}
  \label{fig:x5}
\end{figure}

We expected per-flow fair-queuing to neutralize two-conn's advantage.
The result is the opposite, and it is the empirical face of barrier B3.
Introducing FqCoDel widens two-conn's lead (Fig.~\ref{fig:x5}):
at $K{=}4$ background flows, two-conn's critical-class p50 improves from
154\,ms to 42\,ms ($-73$\%) while CATS-over-TCP's is unchanged or slightly
worse (280\,ms to 333\,ms, $+19$\%); at p99, two-conn (385\,$\to$\,312\,ms)
stays roughly $4{\times}$ ahead of CATS-over-TCP (1{,}569\,$\to$\,1{,}452\,ms)
under both queue disciplines.
The mechanism is exactly B3: FqCoDel isolates per flow, so two-conn's
two five-tuples land in two separate fair queues, giving its critical flow
per-flow isolation, while CATS-over-TCP's single five-tuple keeps
both classes in one queue where fair-queuing cannot tell them apart.
Meanwhile the bandwidth picture is unchanged: two-conn's Jain index stays
0.84 (thin-flow artifact; the two connections together take one aggregate share) and CATS-over-TCP's is $\approx$1.0.
On a managed link, the case for CATS-over-TCP rests entirely on
fairness and non-starvation, not on the tail; no network QoS,
FqCoDel or L4S, changes that for a single flow.

\subsection{Handover: The Single-Connection Envelope, and Why QUIC}
\label{sec:eval:handover}

\begin{figure*}[t]
  \centering
  \subfloat[RTT sweep: P0 p99 tail and completeness-failure rate
    vs RTT (sub-panels (i)\,/\,(ii)).]
    {\includegraphics[width=\textwidth]{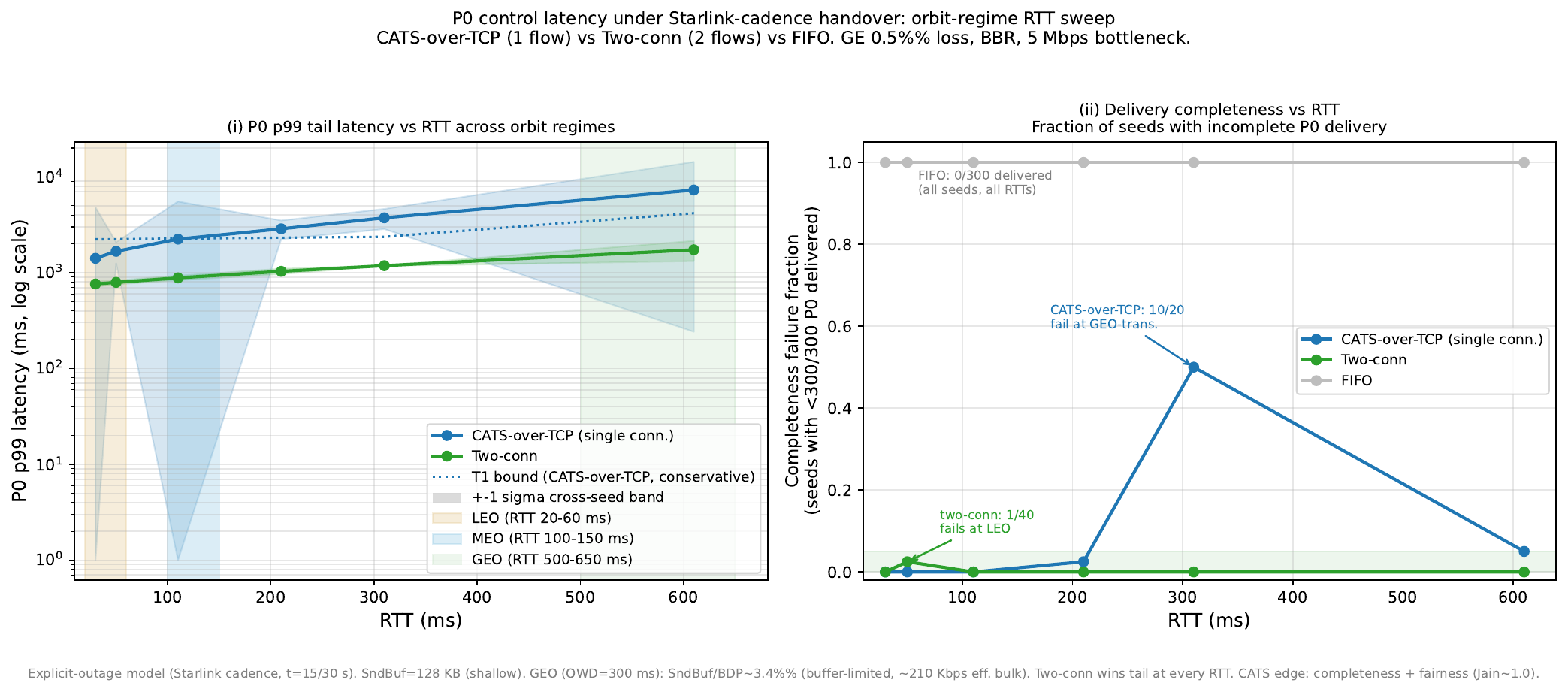}}
  \\[2pt]
  \subfloat[LEO/MEO/GEO anchor CCDFs (OWD\,=\,20/50/300\,ms, 40/20/20 seeds);
    CCDF of per-seed p99 across seeds.]
    {\includegraphics[width=\textwidth]{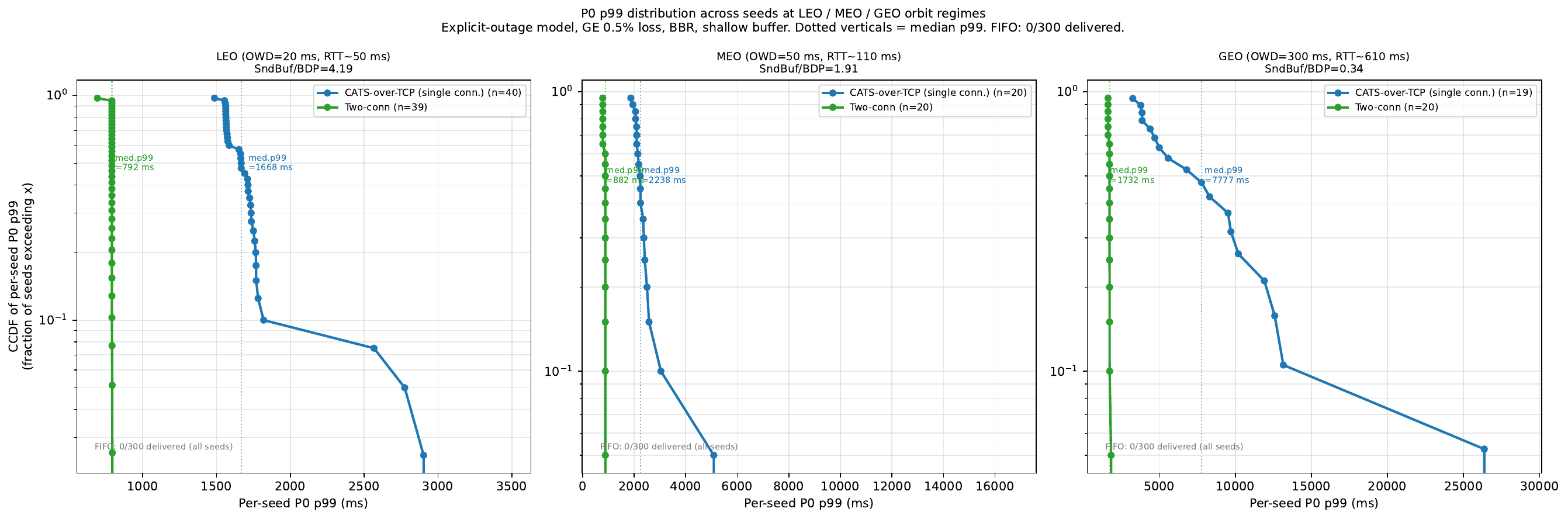}}
  \caption{RTT sweep across orbit regimes (Gilbert-Elliott 0.5\%
    loss, dynamic handover, 20 seeds each OWD point).
    Two connections win the critical-class tail at every RTT, by a
    growing margin (1.85$\times$ at RTT~$\approx$30\,ms to 4.49$\times$ at
    RTT~$\approx$610\,ms).
    CATS-over-TCP's completeness edge is LEO-confined and slight (0/40 incomplete
    at OWD=20\,ms; two-conn 1/40); it inverts at OWD=150\,ms
    (10/20 incomplete, $\approx$50\%).
    FIFO: 0/300 delivered at every RTT.
    Section~\ref{sec:eval:handover} discusses the 40-seed canonical
    (OWD=20\,ms) LEO result.}
  \label{fig:r12}
\end{figure*}

Under link handover, modeled as explicit outages aligned to the Starlink
cadence (Section~\ref{sec:motivation}), we sweep round-trip time across
orbit regimes (LEO through GEO) and report two things:
where single-connection priority works, and where it does not.

Two connections win the critical-class tail at every RTT, and by a
growing margin: 1.85$\times$ at RTT~$\approx$30\,ms (OWD=10\,ms)
to 4.49$\times$ at RTT~$\approx$610\,ms (OWD=300\,ms) (all p99 values
complete-seeds-only medians; see Section~\ref{sec:eval:setup}).
CATS-over-TCP wins the tail nowhere.
Its only advantage is delivery completeness, and that advantage is
confined to the LEO regime and is slight.

At the 40-seed canonical LEO point (OWD=20\,ms):
CATS-over-TCP completes all 40/40 seeds (0/40 incomplete)
where two connections drop one seed (seed 12: 288/300 delivered, 1/40 incomplete);
FIFO delivers 0/300 in every seed.
However, at the same RTT, two-conn wins the tail: median p99 792\,ms
vs.\ 1{,}668\,ms (ratio 2.11$\times$).

This completeness edge inverts as RTT grows:
beyond LEO, single-connection TCP degrades sharply (50\% of seeds
incomplete at OWD=150\,ms, RTT~$\approx$310\,ms), while two
connections stay complete across the range (0/20 incomplete at all OWD points).
FIFO delivers 0/300 at every RTT.

This high-RTT collapse is barrier B2 in its unbounded regime
(Section~\ref{sec:theory:barriers}).
During a handover at high RTT, a single-connection RTO chain drives the
congestion window toward zero, and the credit-based shaper cannot help,
because there are no transmission opportunities for any class;
the failure is one of connection liveness, not of scheduling, and
the per-class non-starvation guarantee is conditioned on the former
(Section~\ref{sec:theory:model}).
Two connections avoid it precisely because a stall on one does not stall
the other, the same flow-count independence that makes them each a separate five-tuple
(Section~\ref{sec:eval:fairness}).
The reading is not that CATS-over-TCP has a handover niche,
but that the sweep characterizes the single-connection envelope: priority
on one TCP connection is viable at LEO round-trip times and fails at
GEO ones.
That is the sharpest motivation for the QUIC escape of Section~\ref{sec:quic}.

\subsection{Loss, CCA-Agnosticism, and One Application}
\label{sec:eval:loss}

Three brief results complete the envelope.
Under increasing loss, a FIFO baseline's critical-class delivery collapses
while CATS-over-TCP holds.
CATS-over-TCP is also agnostic to the congestion controller (results hold under both
BBR and CUBIC) and composes with specialized controllers,
including LEO handover-aware controllers, rather than competing with them.
Finally, one concrete application (not the paper's claim): applying CATS-over-TCP
to web page loading, where the critical class is render-blocking resources,
improves a transfer-completion proxy by up to 78.7\% (worst-case
scenario, reverse-priority enqueue). This is a delivery-time proxy, not
a Web Vitals measurement, and we report it only to ground the mechanism.

\section{CATS over QUIC: The Escape}
\label{sec:quic}

Section~\ref{sec:theory:table} placed CATS-over-QUIC as the only point in
the design space that clears B1 and bounds B2 on a single fair flow.
Here we substantiate that placement with a proof-of-concept, and develop
the realization it requires.

\subsection{Endpoint Self-Isolation}
\label{sec:quic:isolation}

The argument is structural.
QUIC~\cite{rfc9000} carries independent stream sequence spaces, so a loss on the bulk
stream cannot head-of-line-block the critical stream, clearing B1.
Per-stream isolation, a priority-weighted allocation of the aggregate
window, bounds B2 rather than clearing it: the reservation preserves the
critical class's share of the window once a bulk loss cuts it, but it
cannot prevent the cut, because the cut is what keeps the connection one
congestion-fair flow.
Once the endpoint isolates the critical class from the sequence-space
effect this way, it no longer needs the network for that isolation, making
B3 moot from above, not
because QUIC streams are network-visible (they are not: one encrypted
five-tuple), but because no network-layer isolation is required.
All of this holds on one connection, one flow's congestion-fair
share.

\subsection{Proof-of-Concept}
\label{sec:quic:poc}

\begin{figure}[t]
  \centering
  \includegraphics[width=\columnwidth]{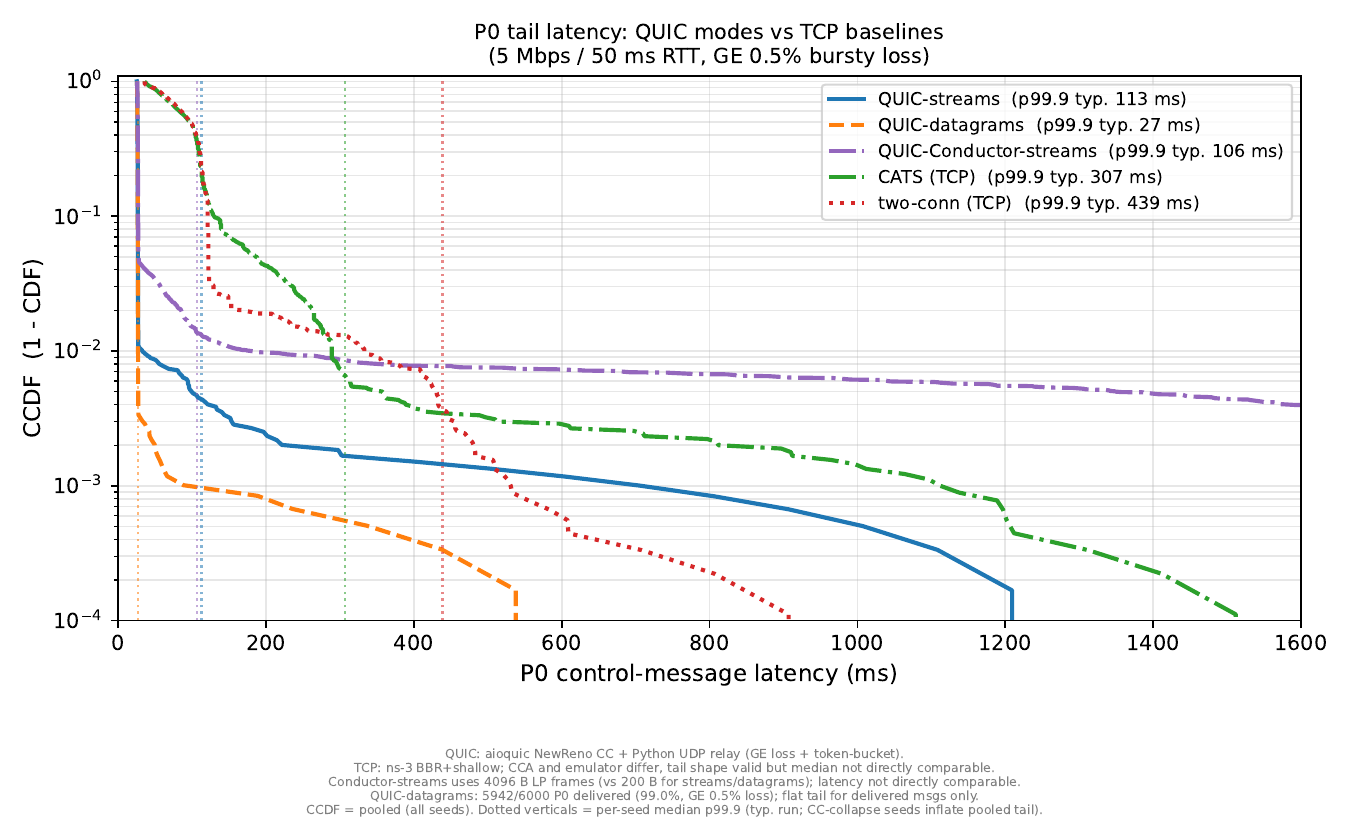}
  \caption{QUIC PoC: P0 critical-class latency CCDF (aioquic 1.3.0,
    GE 0.5\% loss, 20 seeds, NewReno CC).
    Datagrams: flat tail at $\approx$27\,ms p99.9 in 19/20 seeds;
    seed 6 (CC collapse): 638\,ms.
    Streams: p99.9 median 112.6\,ms, above loss-only bound (102.7\,ms)
    by $\Delta_{cc} \approx 25$\,ms (B2, BBR).
    TCP baseline (single-connection CATS-over-TCP): per-seed-median p99.9~$\approx$307\,ms
    (loss-only, no competing flow; Section~\ref{sec:eval:fairness}'s
    4.7$\times$/4--8$\times$ two-connection tail advantage arises under
    cross-traffic contention).
    B2's CC-collapse case visible in seed 6 across all modes.}
  \label{fig:quic}
\end{figure}

\textbf{Setup:}
aioquic 1.3.0, a userspace UDP relay injecting Gilbert-Elliott 0.5\% bursty
loss (no kernel privileges required), 20 seeds, three modes:
(a)~\emph{streams}: P0 on one stream, P4 bulk on another;
(b)~\emph{datagrams}: P0 as QUIC DATAGRAM frames~\cite{rfc9221};
(c)~\emph{conductor-streams}: Conductor-paced streams with larger frames.

The structural bypass holds (Fig.~\ref{fig:quic}).
Independent streams remove head-of-line blocking, and the datagram tail is
flat: critical-class p99.9 is $\approx$27\,ms (median over 19/20
seeds) against $\approx$307\,ms over TCP (per-seed-median p99.9; cf.\ pooled p99.9 = 1{,}112\,ms in \S\ref{sec:eval:single}).
This result is a structural B1-bypass demonstration in which datagrams trade
reliability for the flat tail (5{,}942/6{,}000 messages delivered across
20 seeds, 99.0\%); must-arrive critical traffic uses the reliable streams
mode (p99.9 median 112.6\,ms, bounded by $\Delta_{cc}$ until per-stream isolation).
The effect is not an artifact of the congestion controller:
re-running under both NewReno and CUBIC yields the same head-of-line-free
shape, so the bypass is structural, not CCA-specific.

One result matters more than the flat tail: the case where it \emph{breaks}.
In 1 of 20 seeds (seed 6), a connection-level congestion collapse
(window driven to near-zero for $\approx$600\,ms) stalls even
independent datagrams (638\,ms).
Independent streams bypass per-stream flow control but not
connection-level congestion control. This is barrier B2 in its rare,
unbounded regime (Section~\ref{sec:theory:barriers}), and it identifies
the precise requirement for the escape to be complete: \emph{per-stream
isolation from connection-level congestion collapse}, which today's
QUIC does not provide directly.

The streams p99.9 (median 112.6\,ms) slightly exceeds the loss-only bound
(102.7\,ms) by $\Delta_{cc} \approx 25$\,ms; the CC-corrected bound
($\approx$128\,ms) holds at the median, confirming Lemma~\ref{lem:b2}.

We are explicit about the PoC's limits: the relay is userspace, not a
kernel emulator; aioquic is a reference stack, not production; and since
aioquic's controller (NewReno) differs from the ns-3 baseline (BBR),
we compare tail shape and structural effects, not absolute latency.

\subsection{The Realization: Priority-Weighted Coupled Congestion Control}
\label{sec:quic:realization}

The requirement from Section~\ref{sec:quic:poc}, per-stream isolation from
connection-level congestion, must be met without the extra-flow overhead
of parallel connections.
Independence per stream is therefore the wrong primitive.

The right primitive is hierarchical, coupled congestion control.
One connection-level controller estimates the aggregate fair sending
rate, keeping the connection one flow's share at the bottleneck, and a
priority-weighted allocator partitions that aggregate window across priority
buckets, charging congestion-induced reductions to low-priority buckets
first while the critical class holds a protected reservation.
This decouples the two concerns cleanly: fairness is handled once, at the
aggregate; isolation is handled by the within-connection partition.
It is one fair flow, internally priority-scheduled.

MPTCP's coupled congestion control~\cite{rfc6356} performs exactly this
aggregate-fairness trick across subflows; our design applies the same coupling
across priority buckets.
And the gap is explicit in the standards: RFC~9218~\cite{rfc9218} governs
scheduling but deliberately leaves congestion control untouched,
so priority-aware CC is open territory.

Even perfect partitioning does not give the critical class zero
latency: it remains bounded below by $D_0$ of Theorem~\ref{thm:bound},
carried into QUIC, not eliminated.

Three open directions follow from this boundary.
One is the priority-weighted aggregate controller itself: whether it
can remain one-flow-fair while still isolating the critical class under
adversarial loss patterns designed to defeat the reservation.
Another is extending the same coupling across the multiple paths of a
multi-orbit (LEO + MEO + GEO) terminal, mapping semantic priority to path
(Section~\ref{sec:conclusion}).
A third is the reliability tradeoff underneath the
primitive choice itself. Datagrams buy the flat tail only by dropping
messages (99.0\% delivered in the PoC, Section~\ref{sec:quic:poc}), while
reliable streams keep every message but re-expose B2's coupling. The
coupled controller of this section does not by itself decide which
primitive a class should use; realizing it fully requires jointly
assigning each class its primitive (reliable stream or unreliable
datagram), its urgency~\cite{rfc9218}, and its allocator reservation,
since the three interact.

No free lunch. Tail isolation for the critical class requires a private congestion window.
Two connections buy exactly that, and pay for it with a second fair share.
On one flow, the critical class remains exposed to the aggregate
controller's multiplicative decrease, because that decrease is what makes
the connection one fair flow in the first place.
The escape clears B1 and bounds B2; clearing B2 outright would mean the
connection ceasing to behave as a single congestion-fair flow.
That boundary, not its absence, is the result: it is what separates the
escape from a second connection wearing QUIC's clothing.

\section{Related Work}
\label{sec:related}

\textbf{Priority-aware QUIC congestion control:}
PCC Priority~\cite{pcc_priority} folds stream priorities into a
performance-oriented CC utility function and proves proportional-bandwidth
allocation by priority.
CATS differs in objective: we target latency isolation, a provable
bounded per-class delay, and a protected critical-class reservation
(deadline-determinism), rather than bandwidth proportionality.
Stream-scheduling work (FStream~\cite{fstream}, XLINK~\cite{xlink},
MS-HS~\cite{ms_hs}) arbitrates which stream sends, not how the congestion
window is partitioned; the window-allocation question of
Section~\ref{sec:quic:realization}, with its credit-based-shaper bound,
is the gap they leave open.

\textbf{LEO/NTN transport:}
the community has converged on handover-aware congestion control,
freezing the congestion window across the periodic handover~\cite{leo_mobility,startcp,starquic,satcp}.
CATS is orthogonal: it arbitrates priority within a connection,
while those repair the controller's reaction to a handover.
Closest to our multipath direction is mobility-aware multipath-QUIC CC
for integrated terrestrial-satellite networks~\cite{mm_quic}; we differ
in contributing a semantic priority-to-path mapping with bounded-delay
guarantees.

\textbf{Low-latency and QoS in the network:}
AQM and fair-queuing (CoDel~\cite{rfc8289}, FqCoDel~\cite{rfc8290}) and L4S~\cite{rfc9330,rfc9331,rfc9332}
reduce queuing delay at flow granularity, which is precisely
barrier B3.
Rather than competing with CATS, L4S witnesses B3, confirming the
barrier is a property of all flow-granular network QoS.

\textbf{Deadline-aware and priority transport:}
deadline-aware TCP variants (D2TCP~\cite{d2tcp}, DTP~\cite{dtp}) and
opportunistic-retransmission and proactive-loss-recovery approaches
target latency-sensitive flows but operate across flows or modify the wire
protocol/receiver; CATS is single-connection, sender-side, and
receiver-transparent.
Application-layer priorities (RFC~9218~\cite{rfc9218}) schedule but do not
touch congestion control.
Extensible-kernel-transport work (eTran~\cite{etran}) provides the eBPF
data-path substrate on which a production CATS-over-QUIC (with the
per-stream coupled controller of Section~\ref{sec:quic:realization})
would most naturally be built.

\section{Limitations and Future Work}
\label{sec:limitations}

\textbf{The high-RTT limit of the TCP realization:}
CATS-over-TCP's guarantees are most useful at LEO round-trip times;
the handover sweep (Section~\ref{sec:eval:handover}) shows that at GEO
round-trip times single-connection TCP suffers severe completeness failure
(50\% of seeds at RTT~$\approx$310\,ms), a connection-liveness collapse the
per-class shaper cannot prevent.
This bounds the deployable envelope of the TCP realization to
lower-RTT links and is a direct argument for the QUIC realization.

\textbf{What the simulator cannot show:}
ns-3 resets transport state for every scheme across a modeled outage,
so the regime where a single connection should beat parallel
connections at handover (preserved congestion state and SACK across
the disruption) is not observable here.
Realizing and measuring that advantage requires connection migration
(MPTCP or QUIC).

\textbf{The status of B3:}
barrier B3 is an architectural and empirically-witnessed statement about
flow-granular network QoS, not a deductive theorem, and we present it
as such; a fully formal treatment over a defined class of work-conserving
per-flow schedulers is open.

\textbf{Unvalidated and unbuilt pieces:}
the DSCP mitigation requires AQM-aware hardware we did not test;
by B3 it cannot help a single connection regardless.
The per-stream coupled congestion controller of
Section~\ref{sec:quic:realization} is specified but not built:
a production CATS-over-QUIC in quiche or mvfst, with per-stream
isolation atop a single aggregate fair controller, is the central
next system.

\section{Conclusion}
\label{sec:conclusion}

We asked what a single commodity connection can guarantee for
mixed-criticality traffic, and what it fundamentally cannot.
The answer is a clean separation.
CATS-over-TCP, a sender-side, receiver-transparent, just-in-time priority scheme
over commodity TCP, delivers the one combination its alternatives cannot:
deterministic non-starvation, single-flow fairness, and a provable bound
on per-class delay.
But it is not a tail-latency mechanism, and three structural barriers show
why: the in-order sequence space, the shared congestion window, and the
per-flow granularity of network QoS, which respectively explain the
critical-class tail under loss, the coupling under contention, and why
fair-queuing and L4S cannot help.
Two parallel connections clear all three barriers but require an
additional five-tuple and are bandwidth-fair only when the critical class is
light; CATS-over-QUIC clears B1 and bounds B2 by self-isolating at the
endpoint (independent sequence spaces with per-stream isolation under aggregate-coupled congestion control), mooting B3, on a
single fair flow, and a proof-of-concept confirms the structural bypass
while pinning the precise requirement: per-stream isolation under
aggregate-coupled congestion control, realized without sacrificing single-flow fairness.

That escape is also a program.
The realization it calls for is per-stream, priority-weighted coupled
congestion control that preserves single-flow fairness while isolating
the critical class. Extended across the multiple paths of a multi-orbit
(LEO + MEO + GEO) terminal, it is the natural next step for deterministic
mixed-criticality transport on exactly the satellite and non-terrestrial
links where it matters most.

\bibliographystyle{IEEEtran}
\bibliography{refs}

@inproceedings{cats2025icic,
  author    = {Rizvi, Syed Muhammad Aqdas},
  title     = {{A Case for CATS: A Conductor-driven Asymmetric Transport Scheme for Semantic Prioritization}},
  booktitle = {2025 6th International Conference on Innovative Computing (ICIC)},
  year      = {2025},
  publisher = {IEEE},
  doi       = {10.1109/ICIC68258.2025.11413235},
  url = {https://doi.org/10.1109/ICIC68258.2025.11413235}
}

@misc{cats2026arxiv,
      title={{A Case for CATS: A Conductor-driven Asymmetric Transport Scheme for Semantic Prioritization}}, 
      author={Syed Muhammad Aqdas Rizvi},
      year={2026},
      eprint={2604.16913},
      archivePrefix={arXiv},
      primaryClass={cs.AI},
      url={https://doi.org/10.48550/arxiv.2603.13945},
      note= {Extended version of \cite{cats2025icic}}, 
}

@ARTICLE{pcc_priority,
  author={Chen, Zhuoyue and Cai, Kechao and Zhang, Jinbei and Zhu, Xiangwei},
  journal={IEEE Networking Letters}, 
  title={{PCC Priority: A Priority-Aware Bandwidth Allocation Framework for QUIC}}, 
  year={2023},
  volume={5},
  number={4},
  pages={279-283},
  doi={10.1109/LNET.2023.3269054},
  url = {https://doi.org/10.1109/LNET.2023.3269054}
  }

@INPROCEEDINGS{fstream,
  author={Shi, Xiang and Wang, Lin and Zhang, Fa and Liu, Zhiyong},
  booktitle={2019 IEEE 25th International Conference on Parallel and Distributed Systems (ICPADS)}, 
  title={{FStream: Flexible Stream Scheduling and Prioritizing in Multipath-QUIC}}, 
  year={2019},
  volume={},
  number={},
  pages={921-924},
  doi={10.1109/ICPADS47876.2019.00136},
  url = {https://doi.org/10.1109/ICPADS47876.2019.00136}
}

@inproceedings{xlink,
author = {Zheng, Zhilong and Ma, Yunfei and Liu, Yanmei and Yang, Furong and Li, Zhenyu and Zhang, Yuanbo and Zhang, Jiuhai and Shi, Wei and Chen, Wentao and Li, Ding and An, Qing and Hong, Hai and Liu, Hongqiang Harry and Zhang, Ming},
title = {{XLINK: QoE-driven multi-path QUIC transport in large-scale video services}},
year = {2021},
isbn = {9781450383837},
publisher = {Association for Computing Machinery},
address = {New York, NY, USA},
url = {https://doi.org/10.1145/3452296.3472893},
doi = {10.1145/3452296.3472893},
booktitle = {Proceedings of the 2021 ACM SIGCOMM 2021 Conference},
pages = {418-432},
numpages = {15},
location = {Virtual Event, USA},
series = {SIGCOMM '21}
}

@INPROCEEDINGS{ms_hs,
  author={Liang, Xiangbin and Zhao, Baokang and Peng, Wei and Wang, Tianshu},
  booktitle={2022 10th International Conference on Information Systems and Computing Technology (ISCTech)}, 
  title={{Towards Effective Multipath Scheduling with Multipath QUIC in Heterogeneous Paths}}, 
  year={2022},
  volume={},
  number={},
  pages={472-479},
  doi={10.1109/ISCTech58360.2022.00079},
  url = {https://doi.org/10.1109/ISCTech58360.2022.00079}
  }

@inproceedings{startcp,
author = {Jiang, Li and Zhang, Yihang and Hu, Yannan and Cui, Yong and Zhang, Xinggong},
title = {{StarTCP: Handover-aware Transport Protocol for Starlink}},
year = {2024},
isbn = {9798400717581},
publisher = {Association for Computing Machinery},
address = {New York, NY, USA},
url = {https://doi.org/10.1145/3663408.3665803},
doi = {10.1145/3663408.3665803},
booktitle = {Proceedings of the 8th Asia-Pacific Workshop on Networking},
pages = {169--170},
numpages = {2},
location = {Sydney, Australia},
series = {APNet '24}
}

@inproceedings{starquic,
author = {Kamel, Victor and Zhao, Jinwei and Li, Daoping and Pan, Jianping},
title = {{StarQUIC: Tuning Congestion Control Algorithms for QUIC over LEO Satellite Networks}},
year = {2024},
isbn = {9798400712807},
publisher = {Association for Computing Machinery},
address = {New York, NY, USA},
url = {https://doi.org/10.1145/3697253.3697271},
doi = {10.1145/3697253.3697271},
booktitle = {Proceedings of the 2nd International Workshop on LEO Networking and Communication},
pages = {43-48},
numpages = {6},
location = {Washington, DC, USA},
series = {LEO-NET '24}
}

@INPROCEEDINGS{satcp,
  author={Cao, Xuyang and Zhang, Xinyu},
  booktitle={IEEE INFOCOM 2023 - IEEE Conference on Computer Communications}, 
  title={{SaTCP: Link-Layer Informed TCP Adaptation for Highly Dynamic LEO Satellite Networks}}, 
  year={2023},
  volume={},
  number={},
  pages={1-10},
  doi={10.1109/INFOCOM53939.2023.10228914},
  url = {https://doi.org/10.1109/INFOCOM53939.2023.10228914}
  }

@inproceedings{leo_mobility,
author = {Lai, Zeqi and Li, Zonglun and Wu, Qian and Li, Hewu and Liu, Weisen and Liu, Yijie and Xie, Xin and Li, Yuanjie and Liu, Jun},
title = {{Mind the Misleading Effects of LEO Mobility on End-to-End Congestion Control}},
year = {2024},
isbn = {9798400712722},
publisher = {Association for Computing Machinery},
address = {New York, NY, USA},
url = {https://doi.org/10.1145/3696348.3696867},
doi = {10.1145/3696348.3696867},
booktitle = {Proceedings of the 23rd ACM Workshop on Hot Topics in Networks},
pages = {34-42},
numpages = {9},
location = {Irvine, CA, USA},
series = {HotNets '24}
}

@ARTICLE{mm_quic,
  author={Yang, Wenjun and Cai, Lin and Shu, Shengjie and Pan, Jianping},
  journal={IEEE Transactions on Mobile Computing}, 
  title={{Mobility-Aware Congestion Control for Multipath QUIC in Integrated Terrestrial Satellite Networks}}, 
  year={2024},
  volume={23},
  number={12},
  pages={11620-11634},
  doi={10.1109/TMC.2024.3397164},
  url = {https://doi.org/10.1109/TMC.2024.3397164}
  }

@misc{rfc9330,
    series =    {Request for Comments},
    number =    9330,
    howpublished =  {RFC 9330},
    publisher = {RFC Editor},
    doi =       {10.17487/RFC9330},
    url =       {https://www.rfc-editor.org/info/rfc9330},
    author =    {Bob Briscoe and Koen De Schepper and Marcelo Bagnulo and Greg White},
    title =     {{Low Latency, Low Loss, and Scalable Throughput (L4S) Internet Service: Architecture}},
    pagetotal = 36,
    year =      2023,
    month =     jan,
}

@misc{rfc9331,
    series =    {Request for Comments},
    number =    9331,
    howpublished =  {RFC 9331},
    publisher = {RFC Editor},
    doi =       {10.17487/RFC9331},
    url =       {https://www.rfc-editor.org/info/rfc9331},
    author =    {Koen De Schepper and Bob Briscoe},
    title =     {{The Explicit Congestion Notification (ECN) Protocol for Low Latency, Low Loss, and Scalable Throughput (L4S)}},
    pagetotal = 52,
    year =      2023,
    month =     jan,
}

@misc{rfc9332,
    series =    {Request for Comments},
    number =    9332,
    howpublished =  {RFC 9332},
    publisher = {RFC Editor},
    doi =       {10.17487/RFC9332},
    url =       {https://www.rfc-editor.org/info/rfc9332},
    author =    {Koen De Schepper and Bob Briscoe and Greg White},
    title =     {{Dual-Queue Coupled Active Queue Management (AQM) for Low Latency, Low Loss, and Scalable Throughput (L4S)}},
    pagetotal = 52,
    year =      2023,
    month =     jan,
}

@misc{rfc9221,
    series =    {Request for Comments},
    number =    9221,
    howpublished =  {RFC 9221},
    publisher = {RFC Editor},
    doi =       {10.17487/RFC9221},
    url =       {https://www.rfc-editor.org/info/rfc9221},
    author =    {Tommy Pauly and Eric Kinnear and David Schinazi},
    title =     {{An Unreliable Datagram Extension to QUIC}},
    pagetotal = 9,
    year =      2022,
    month =     mar,
}

@misc{rfc9218,
    series =    {Request for Comments},
    number =    9218,
    howpublished =  {RFC 9218},
    publisher = {RFC Editor},
    doi =       {10.17487/RFC9218},
    url =       {https://www.rfc-editor.org/info/rfc9218},
    author =    {Kazuho Oku and Lucas Pardue},
    title =     {{Extensible Prioritization Scheme for HTTP}},
    pagetotal = 21,
    year =      2022,
    month =     jun,
}

@misc{rfc9000,
    series =    {Request for Comments},
    number =    9000,
    howpublished =  {RFC 9000},
    publisher = {RFC Editor},
    doi =       {10.17487/RFC9000},
    url =       {https://www.rfc-editor.org/info/rfc9000},
    author =    {Jana Iyengar and Martin Thomson},
    title =     {{QUIC: A UDP-Based Multiplexed and Secure Transport}},
    pagetotal = 151,
    year =      2021,
    month =     may,
}

@misc{rfc6356,
    series =    {Request for Comments},
    number =    6356,
    howpublished =  {RFC 6356},
    publisher = {RFC Editor},
    doi =       {10.17487/RFC6356},
    url =       {https://www.rfc-editor.org/info/rfc6356},
    author =    {Costin Raiciu and Mark J. Handley and Damon Wischik},
    title =     {{Coupled Congestion Control for Multipath Transport Protocols}},
    pagetotal = 12,
    year =      2011,
    month =     oct,
}

@inproceedings{d2tcp,
author = {Vamanan, Balajee and Hasan, Jahangir and Vijaykumar, T.N.},
title = {Deadline-aware datacenter tcp (D2TCP)},
year = {2012},
isbn = {9781450314190},
publisher = {Association for Computing Machinery},
address = {New York, NY, USA},
url = {https://doi.org/10.1145/2342356.2342388},
doi = {10.1145/2342356.2342388},
booktitle = {Proceedings of the ACM SIGCOMM 2012 Conference on Applications, Technologies, Architectures, and Protocols for Computer Communication},
pages = {115-126},
numpages = {12},
location = {Helsinki, Finland},
series = {SIGCOMM '12}
}

@inproceedings{dtp,
author = {Shi, Hang and Cui, Yong and Qian, Feng and Hu, Yuming},
title = {DTP: Deadline-aware Transport Protocol},
year = {2019},
isbn = {9781450376358},
publisher = {Association for Computing Machinery},
address = {New York, NY, USA},
url = {https://doi.org/10.1145/3343180.3343191},
doi = {10.1145/3343180.3343191},
booktitle = {Proceedings of the 3rd Asia-Pacific Workshop on Networking},
pages = {1-7},
numpages = {7},
location = {Beijing, China},
series = {APNet '19}
}

@inproceedings{etran,
author = {Chen, Zhongjie and Meng, Qingkai and Lao, ChonLam and Liu, Yifan and Ren, Fengyuan and Yu, Minlan and Zhou, Yang},
title = {eTran: extensible kernel transport with eBPF},
year = {2025},
isbn = {978-1-939133-46-5},
publisher = {USENIX Association},
address = {USA},
booktitle = {Proceedings of the 22nd USENIX Symposium on Networked Systems Design and Implementation},
articleno = {22},
numpages = {19},
location = {Philadelphia, PA, USA},
series = {NSDI '25},
doi = {10.5555/3767955.3767977},
url = {https://doi.org/10.5555/3767955.3767977}
}

@article{bbr,
author = {Cardwell, Neal and Cheng, Yuchung and Gunn, C. Stephen and Yeganeh, Soheil Hassas and Van Jacobson},
title = {BBR: congestion-based congestion control},
year = {2017},
issue_date = {February 2017},
publisher = {Association for Computing Machinery},
address = {New York, NY, USA},
volume = {60},
number = {2},
issn = {0001-0782},
url = {https://doi.org/10.1145/3009824},
doi = {10.1145/3009824},
journal = {Commun. ACM},
month = jan,
pages = {58-66},
numpages = {9}
}

@misc{rfc9438,
    series =    {Request for Comments},
    number =    9438,
    howpublished =  {RFC 9438},
    publisher = {RFC Editor},
    doi =       {10.17487/RFC9438},
    url =       {https://www.rfc-editor.org/info/rfc9438},
    author =    {Lisong Xu and Sangtae Ha and Injong Rhee and Vidhi Goel and Lars Eggert},
    title =     {{CUBIC for Fast and Long-Distance Networks}},
    pagetotal = 28,
    year =      2023,
    month =     aug,
}

@misc{rfc6582,
    series =    {Request for Comments},
    number =    6582,
    howpublished =  {RFC 6582},
    publisher = {RFC Editor},
    doi =       {10.17487/RFC6582},
    url =       {https://www.rfc-editor.org/info/rfc6582},
    author =    {Andrei Gurtov and Tom Henderson and Sally Floyd and Yoshifumi Nishida},
    title =     {{The NewReno Modification to TCP's Fast Recovery Algorithm}},
    pagetotal = 16,
    year =      2012,
    month =     apr,
}

@misc{rfc8289,
    series =    {Request for Comments},
    number =    8289,
    howpublished =  {RFC 8289},
    publisher = {RFC Editor},
    doi =       {10.17487/RFC8289},
    url =       {https://www.rfc-editor.org/info/rfc8289},
    author =    {Kathleen Nichols and Van Jacobson and Andrew McGregor and Jana Iyengar},
    title =     {{Controlled Delay Active Queue Management}},
    pagetotal = 25,
    year =      2018,
    month =     jan,
}

@misc{rfc8290,
    series =    {Request for Comments},
    number =    8290,
    howpublished =  {RFC 8290},
    publisher = {RFC Editor},
    doi =       {10.17487/RFC8290},
    url =       {https://www.rfc-editor.org/info/rfc8290},
    author =    {Toke H{\o}iland-J{\o}rgensen and Paul McKenney and Dave T{\"a}ht and Jim Gettys and Eric Dumazet},
    title =     {{The Flow Queue CoDel Packet Scheduler and Active Queue Management Algorithm}},
    pagetotal = 25,
    year =      2018,
    month =     jan,
}

@misc{jain_fairness,
author = {Jain, R. and Chiu, D. and Hawe, W.},
title = {{A Quantitative Measure Of Fairness And Discrimination For Resource Allocation In Shared Computer Systems}},
year = {1998},
eprint = {cs/9809099},
archivePrefix = {arXiv},
url = {https://arxiv.org/abs/cs/9809099},
note = {Originally DEC Research Report TR-301, September 1984}
}

\end{document}